# Evidence for Radiogenic Sulfur-32 in Type AB Presolar Silicon Carbide Grains?

Wataru Fujiya[1,2], Peter Hoppe[1], Ernst Zinner[3], Marco Pignatari[4,5], and Falk Herwig[5,6]

[1]Max Planck Institute for Chemistry, Hahn-Meitner-Weg 1, 55128 Mainz, Germany
(wataru.fujiya@mpic.de, peter.hoppe@mpic.de)

[2]Department of Earth and Planetary Science, University of Tokyo, 7-3-1 Hongo, Bunkyo-ku, 113-0033 Tokyo, Japan

[3]Laboratory for Space Sciences and Physics Dept., Campus Box 1105, Washington University, St. Louis, MO 63130, USA

(ekz@wustl.edu)

[4]Department of Physics, University of Basel, Klingelbergstrasse 82, CH-4056 Basel, Switzerland

(mpignatari@gmail.com)

[5]NuGrid collaboration, http://www.nugridstars.org

[6]Department of Physics & Astronomy, University of Victoria, Canada, P.O. Box 3055, Victoria, BC V8W 3P6, Canada

(fherwig@uvic.ca)

ABSTRACT

We report C, Si, and S isotope measurements on 34 presolar silicon carbide grains of Type AB, characterized by $^{12}C/^{13}C < 10$. Nitrogen, Mg-Al-, and Ca-Ti-isotopic compositions were measured on a subset of these grains. Three grains show large $^{32}S$ excesses, a signature that has been previously observed for grains from supernovae (SNe). Enrichments in $^{32}S$ may be due to contributions from the Si/S zone and the result of S molecule chemistry in still unmixed SN ejecta or due to incorporation of radioactive $^{32}Si$ from C-rich explosive He shell ejecta. However, a SN origin remains unlikely for the three AB grains considered here, because of missing evidence for $^{44}Ti$, relatively low $^{26}Al/^{27}Al$ ratios (a few times $10^{-3}$), and radiogenic $^{32}S$ along with low $^{12}C/^{13}C$ ratios. Instead, we show that born-again asymptotic giant branch (AGB) stars that have undergone a very-late thermal pulse (VLTP), known to have low $^{12}C/^{13}C$ ratios and enhanced abundances of the light s-process elements, can produce $^{32}Si$, which makes such stars attractive sources for AB grains with $^{32}S$ excesses. This lends support to the proposal that at least some AB grains originate from born-again AGB stars, although uncertainties in the born-again AGB star models and possible variations of initial S-isotopic compositions in the parent stars of AB grains make it difficult to draw a definitive conclusion.

# INTRODUCTION

Small quantities of stardust are found in primitive meteorites, interplanetary dust particles, and cometary matter (Zinner 2013). These so-called presolar grains condensed in the winds of low-mass asymptotic giant branch (AGB) stars and in the ejecta of stellar explosions. Presolar grains carry huge isotopic anomalies for various elements, which reflect nucleosynthetic and mixing processes in their parent stars. They survived the passage through the interstellar medium (ISM) and solar system formation largely intact, making them unique samples of stellar material that can be analyzed in detail in the laboratory.

Silicon carbide (SiC) is the best-studied presolar mineral. The majority of presolar SiC grains (~90 %) belong to the so-called "mainstream" (MS) category, characterized by $^{12}C/^{13}C$ ratios of 10 – 100, $^{14}N/^{15}N$ ratios with a wide distribution but, on average, higher than solar, and Si-isotopic compositions that in a Si-three isotope representation plot along a line with slope ~ 1.3 (the Si "MS line"; Zinner et al. 2006). The isotopic compositions of C and Si and, more importantly, of heavy trace elements contained in MS grains indicate that they originate from 1 – 3 $M_\odot$ AGB stars of close-to-solar metallicity (Lugaro et al. 2003).

Silicon carbide grains from Type II supernovae (SNIIe), the X and C grains, are rare. Note that previously the C grains have also been named "unusual" or U/C grains. Here, we define C grains as having $\delta^{29,30}Si$ >300 ‰ ($\delta^i Si = [(^iSi/^{28}Si)_{grain}/[(^iSi/^{28}Si)_\odot - 1] \times$ 1000) and $^{12}C/^{13}C > 10$. The X grains have isotopically light Si, i.e., large excesses in $^{28}Si$ ($\delta^{29,30}Si < -100$ ‰;), isotopically light or heavy C, heavy N, and high inferred $^{26}Al/^{27}Al$ ratios of 0.1 – 1 (e.g., Hoppe et al 2000; Lin et al. 2010). Observed $^{44}Ca$ excesses can be satisfactorily explained only by the decay of $^{44}Ti$, which is synthesized in significant amounts only in SNIIe (Woosley et al. 1973; Timmes et al. 1996; Pignatari et al., 2013a), providing definitive proof of the SNII origin of X grains (Nittler et al. 1996; Hoppe et al. 1996). Except for isotopically heavy Si, C grains exhibit the same isotopic signatures as X grains (Hoppe et al. 2012), including $^{32}S$ excesses. In X grains, $^{32}S$ could have originated either from the Si/S zone (e.g., Rauscher et al. 2002) or from the bottom of the He/C zone experiencing explosive He burning at high shock velocities (Pignatari et al. 2013a). In C grains, $^{32}$S results instead from the decay of radioactive $^{32}$Si (half-life 153 yr; Pignatari et al. 2013b).

SiC grains of Type AB have $^{12}$C/$^{13}$C < 10 (Amari et al. 2001). They account for ~4% of the presolar SiC population. N-isotopic compositions are highly variable, with $^{14}$N/$^{15}$N ratios ranging from 30 to 10,000. While the AB grains have Si-isotopic compositions similar to those of MS grains, they tend to have higher $^{26}$Al/$^{27}$Al ratios of typically $10^{-3}$ to $10^{-2}$ (Amari et al. 2001). Unlike the origin of MS grains, that of AB grains is still a matter of debate. The proposed stellar sources are J-type carbon stars (Lambert et al. 1986) and born-again AGB stars, such as Sakurai's object (Asplund et al. 1999). J-type carbon stars, which have low $^{12}$C/$^{13}$C ratios, are viable candidate sources of AB grains without enrichments in s-process elements (Hedrosa et al. 2013). Unfortunately, the nature of J-type carbon stars is still not well understood; but a proposed scenario involves stellar mergers recently investigated by Zhang & Jeffery (2013). Born-again AGB stars might be the source of AB grains with enhancements of s-process elements, in line with astronomical observations of Sakurai's object (Asplund et al. 1999). These stars experience a very-late thermal pulse (VLTP) where convective-reactive nucleosynthesis occurs and unprocessed, H-rich material is convectively mixed into the He-burning zone (Herwig et al. 2011). For the outer He intershell, which extends almost up to the stellar surface, Herwig et al. (2011) predict a $^{12}$C/$^{13}$C ratio of <10 and high abundances of the first-peak s-process elements such as Sr, Y and Zr. The latter result from high neutron densities (up to a few $10^{15}$ cm$^{-3}$) in the He intershell, produced by α-capture on $^{13}$C (i-process; Cowan & Rose 1977). Dust around Sakurai's object is dominated by amorphous carbon (Chesneau et al. 2009). SiC dust is below the detection limit, but has been identified in spectra from post-AGB stars (Molster et al. 2002).

Amari et al. (2001) excluded SNIIe as possible sources of AB grains. However, later work by Savina et al. (2003, 2007) revealed two AB grains whose Mo- and Ru-isotopic compositions are suggestive of a SN origin. Therefore, SNIIe could be a minor source of AB grains. SiC grains with accepted SN origins (i.e., X and C grains) show characteristic Si-isotopic anomalies, which are not seen in AB grains. Other isotopic fingerprints of SN grains, e.g., high $^{26}$Al/$^{27}$Al (>0.1) and/or evidence for $^{44}$Ti must thus be used to identify possible AB grains from SNIIe.

In the present work, we report C-, N-, Mg-Al-, Si-, S- and Ca-Ti-isotopic measurements on 34 AB grains. These measurements were motivated by the desire to obtain further constraints on the origin of AB grains, in particular to investigate whether SNIIe might have contributed to the population of AB grains and to discuss the proposed origin from born-again AGB stars.

## EXPERIMENTAL

SiC grains were extracted from a 30g sample of the Murchison CM2 meteorite (Besmehn & Hoppe 2003), by using a technique similar to that of Amari et al. (1994). Thousands of SiC grains were dispersed on several clean Au foils, one of which was used for the present study.

The search for SiC AB grains was performed by applying the automatic grain mode analysis with the Cameca NanoSIMS 50 ion probe at Washington University (Gyngard et al. 2010b). Carbon- and Si-isotopic measurements of SiC grains were carried out in three steps: First, we obtained ion images of $^{12}C^-$, $^{13}C^-$, $^{28}Si^-$, $^{29}Si^-$, and $^{30}Si^-$ in multi-collection mode, produced by rastering (256 × 256 pixels, 5000 μs/pixel) a focused primary $Cs^+$ ion beam (~100 nm, ~1 pA) over 20 × 20 $\mu m^2$-sized areas on the sample surface. Second, SiC grains were automatically identified from the $^{28}Si$ ion image, and then each identified SiC grain was analyzed by integrating ion intensities with integration times adjusted according to grain size. Finally, the sample stage was moved to an adjacent area and the above steps were repeated. We acquired C and Si isotope data for ~2300 presolar SiC grains from 464 areas. 105 AB grains were identified in this way. Among those we selected 34 grains with Si-isotopic compositions away from the Si MS line by more than 2σ (Fig. 1) for further analysis.

Subsequent N-, Mg-Al-, S- and Ca-Ti-isotope measurements were carried out with the NanoSIMS 50 ion probe at the Max Plank Institute for Chemistry. Sulfur-isotopic compositions were determined for all 34 grains by acquiring $^{32}S^-$, $^{33}S^-$, and $^{34}S^-$ along with $^{12}C^-$ and $^{13}C^-$ ion images in multi-collection, produced by rastering a $Cs^+$ primary ion beam (~100 nm, ~1pA) over 2 × 2 to 3 × 3 $\mu m^2$ areas (128 × 128 pixels, 15000 µs/pixel, 1 to 6 image planes) covering the AB grains. Subsequently, we carried out Al-Mg and Ca-Ti isotope analyses for 20 and 10 grains, respectively. For the Al-Mg-isotope measurements, $^{24}Mg^+$, $^{25}Mg^+$, $^{26}Mg^+$, $^{27}Al^+$, and $^{28}Si^+$ ion images were obtained in multi-collection by rastering an O⁻ primary ion beam (~300 – 400 nm, ~5 pA) over 2 × 2 to 4 × 4 µm² areas (128 × 128 pixels, 15000 µs/pixel, 4 to 5 image planes). Similarly, $^{28}Si^+$, $^{40}Ca^+$, $^{42}Ca^+$, $^{44}Ca^+$ and $^{48}Ti^+$ ion images were acquired for the Ca-Ti isotope measurements. Finally, N-isotopic compositions were measured on two AB grains by recording $^{12}C^-$, $^{13}C^-$, $^{26}CN^-$, $^{27}CN^-$, and $^{28}Si^-$ ion images in a setup similar to that for the S isotope measurements.

## RESULTS AND DISCUSSION

The isotopic compositions and trace element abundances of 34 AB grains studied here are shown in Table 1. Their Si- and S-isotopic compositions are displayed in Figs. 1 and 2. The Si-isotopic compositions of the AB grains from this study are similar to those from previous studies (Amari et al. 2001). A few grains, however, show relatively large deviations from the Si MS line (AB1, AB27, AB29, AB33, AB34; see Fig. 1). Typical S abundances range from 0.1 to 1 weight percent; in a few cases very high S concentrations are observed (AB13, AB17, AB34), which indicates that AB grains are contaminated to various extent by terrestrial or meteoritic sulfur. Three grains with comparatively low S concentrations and no apparent S contamination show enrichments in $^{32}S$ of >100 ‰ at significance levels of ≥2σ in $^{33}S/^{32}S$ and $^{34}S/^{32}S$ (AB21, AB24) or of ~3σ in $^{34}S/^{32}S$ (AB40) (Fig. 2). Eighteen out of the 20 AB grains analyzed for Mg-Al show excesses in $^{26}Mg$ with inferred $^{26}Al/^{27}Al$ ratios of typically between $10^{-3}$ and $10^{-2}$, consistent with results from previous studies (Amari et al. 2001). One grain (AB34) exhibits a $^{26}Al/^{27}Al$ ratio of $5.7\pm0.8 \times 10^{-2}$, the highest value for AB grains found so far and close to, but still lower than the typical values for X grains. Measurements on some grains (AB10, AB22, AB25, AB36) gave very high Al abundances, which must be caused by Al contamination. No grains show significant $^{42}Ca/^{40}Ca$ and $^{44}Ca/^{40}Ca$ anomalies of >2σ, and therefore there is no clear evidence for extinct $^{44}Ti$.

In the following, we concentrate on the three grains with significant $^{32}$S excesses. On average, these three grains have $\delta^{33}$S ~ –360 ‰ and $\delta^{34}$S ~ –200 ‰. Given the relatively small number of analyzed grains, it is unlikely that the $^{32}$S excesses of these three grains just represent statistical outliers. Neither AGB stars nor novae are expected to produce lower than solar $^{33}$S/$^{32}$S (and $^{34}$S/$^{32}$S for AGB stars) ratios (José et al. 2004; Cristallo et al. 2009). Galactic chemical evolution (GCE) may account for moderate variations of S-isotopic compositions (Kobayashi et al. 2011). But it should be noted that no MS grains show large $^{32}$S excesses as inferred from a study of a few large MS grains (LU and LS separates; Gyngard et al. 2007).

The Rauscher et al. (2002) SNII models predict large enrichments in $^{32}$S for the Si/S zone, dominated by $^{28}$Si and $^{32}$S (Fig. 3). Simple ad-hoc SN mixing models predict $^{32}$S excesses to be accompanied by $^{28}$Si excesses, which are not observed here. If we assume that the $^{32}$S excesses originate from the Si/S zone, we must invoke S-Si fractionation due to sulfur molecule chemistry in the still unmixed SNII ejecta (Hoppe et al. 2012) to explain the Si- and S-isotopic signatures of the three AB grains. If we consider selective mixing of matter from different zones in the 15 M$_\odot$ SNII model of Rauscher et al. (2002) in proportions Si/S:O/Si:He/N:H = 0.0028:0.0094:0.25:1 (mass fractions of whole zones), and assume preferential trapping of S from the Si/S zone by a factor of 20, we find a fairly good match between the observed and predicted isotopic compositions (model A, Table 1). The only mismatch between the grain data and the model prediction exists for the C-isotopic composition; the average $^{12}$C/$^{13}$C ratio of the three grains is ~3 while the predicted value is 17. In SNII models, very low $^{12}$C/$^{13}$C ratios of ~3 are predicted for the He/N zone, but contributions of matter from this zone must be limited because of the observed low $^{26}$Al abundances (the He/N zone is a major production site for $^{26}$Al; Fig. 3).

Recently, Pignatari et al. (2013b) presented an alternative explanation for $^{32}$S excesses in C grains: decay of radioactive $^{32}$Si, produced by n-capture reactions in the C-rich explosive He/C zone. This scenario is attractive because the ad hoc assumption of Si-S fractionation due to S molecule chemistry in SN ejecta can be avoided and because the very large $^{32}$S excess found by Xu et al. (2012) in a C grain, larger than what is predicted for S in the Si/S zone (Fig. 2), can easily be explained. If we follow the approach of Pignatari et al. (2013b), assuming that all $^{33}$S and $^{34}$S in the three AB grains is contamination, $^{32}$Si/$^{28}$Si is calculated from $-0.001 \times \delta S \times {}^{32}$S/$^{28}$Si, where $\delta S$ is the average of $\delta^{33}$S and $\delta^{34}$S. With $\delta S = -250$ ‰ (weighted average of our three AB grains) and $^{32}$S/$^{28}$Si $= 5 \times 10^{-3}$ we obtain $^{32}$Si/$^{28}$Si $= 1.3 \times 10^{-3}$. The model of Pignatari et al. (2013b) predicts $^{32}$Si/$^{28}$Si ratios on the order of $10^{-4}$ to $10^{-3}$, similar to those inferred for SN grains, in certain regions of the C-rich explosive He shell. From the 15 M$_\odot$ SNII model of Rauscher et al. (2002), in which $^{32}$Si is significantly produced in the outer part of the O/C zone (Fig. 3), the average Si-isotopic compositions, and $^{26}$Al/$^{27}$Al and $^{32}$Si/$^{28}$Si ratios can be well reproduced if matter from the Si/S, O/Ne, and O/C (only outer 0.2 M$_\odot$) zones are mixed in proportions 0.0163:0.0136:1 (SN model B, Table 1). However, the $^{32}$Si-rich material from the C-rich layer of Pignatari et al. (2013b) and the mixture in SN model B have high $^{12}$C/$^{13}$C ratios typical of He burning conditions, which is not compatible with the AB grain signatures. Admixture of matter from the $^{13}$C-rich He/N zone is not able to resolve this problem because it would make $^{26}$Al/$^{27}$Al significantly higher than observed. Therefore, in view of no evidence for $^{44}$Ti, the relatively low $^{26}$Al/$^{27}$Al ratios, and especially the low $^{12}$C/$^{13}$C ratios of the grains, a possible SN origin of our three AB grains with $^{32}$S excesses seems unlikely for both stellar models considered, although it cannot be completely excluded due to the large uncertainties affecting those models and present limitations in our understanding of core-collapse SN explosions in general. Note that mixtures in both SN models have C/O $< 1$. It is not favorable for the formation of carbonaceous dust, although Clayton et al. (1999) argued that carbon dust can condense even when C/O $< 1$.

We will thus reconsider born-again AGB stars and investigate whether such stars can produce $^{32}$S excesses. First, we explored the i-process in a 1-zone He intershell with initially high abundances of protons and $^{12}$C, such as they occur in the VLTP in post-AGB star Sakurai's object (Asplund et al. 1999; Werner & Herwig 2006; Herwig et al. 2011). The key ingredient of the i-process is strong proton capture resulting in an efficient production of neutrons due to the $^{13}$C($\alpha$,n)$^{16}$O activation. A 1-zone He intershell, burning at a constant temperature of $2 \times 10^8$ K and $\rho = 10^4$ g/cm$^3$, and an initial abundance of X($^1$H) $= 0.2$, X($^{12}$C) $= 0.5$, and X($^{16}$O) $= 0.35$, with all other species having solar abundances (Asplund 2005), produces a neutron density of N$_n \sim 10^{15}$ cm$^{-3}$, which defines the i-process. This neutron density is much higher than the highest s-process values of N$_n \sim 10^{11\text{-}12}$ cm$^{-3}$, but much lower than in r-process conditions with $N_n > 10^{20}$ cm$^{-3}$. Significant amounts of radioactive $^{32}$Si are predicted to be produced after 1.1 hr, reaching a maximum when large amounts of the first-peak s-process elements (e.g., Sr, Y and Zr) are produced, but before significant production of the second-peak s-process elements (e.g., Ba and La). This matches well the elemental abundances observed in Sakurai's object, which seems not to have experienced any relevant s-process during its AGB phase. Variations in s-process element abundances of AB grains found by Amari et al. (2001) could reflect the fact that AB grains condensed also around post-AGB stars which experienced significant s-processing, producing the second-peak s-process elements before the i-process activation.

Second, we used the one-dimensional, multi-zone model of Herwig et al. (2011) to calculate abundance profiles of C and Si isotopes in the He intershell (Fig. 4). Model RUN106 of Herwig et al. (2011) produces an abundance ratio of the second- to first-peak s-process elements close to that observed in Sakurai's object (Asplund et al. 1999). In the outer He intershell, a $^{32}$Si/$^{28}$Si ratio of ~8 × 10$^{-2}$ is predicted, significantly higher than that inferred for our AB grains. If we assume mixing with unprocessed material of solar composition in a ratio 1:35 by mass (this corresponds to a $^{28}$Si abundance ratio of 1:60), it is possible to reproduce the inferred $^{32}$Si/$^{28}$Si ratio of ~10$^{-3}$ in AB grains. The dilution factor of 35 may be achieved if a likely gradient of i-process products in the outermost zones is considered, as it was observed on the surface of Sakurai's object over a timescale of a few months. However, a proper treatment requires full multi-dimensional hydrodynamic simulations including the energy feedback during H ingestion (Herwig et al. 2011 and references therein), which is beyond the scope of this paper. Along with $^{32}$Si, significant amounts of $^{29}$Si and $^{30}$Si are produced, with higher enrichments in $^{30}$Si than $^{29}$Si. With the assumed dilution of a factor of 35 with isotopically normal material, and $^{29}$Si/$^{28}$Si = 0.16 and $^{30}$Si/$^{28}$Si = 0.35 resulting from the i-process (Fig. 4), we obtain $\delta^{29}$Si = 37 ‰ and $\delta^{30}$Si = 161 ‰. Considering the limitations of the VLTP model used here and the fact that it is based only on one mass and one metallicity, we find the difference between predicted and observed $\delta^{30}$Si acceptable (Table 1). Models of born-again-AGB stars predict $^{12}$C/$^{13}$C ratios of <10 in the outer region of the He intershell along with high C abundances of more than 100 times the solar value. In a 1:35 mixture of He intershell material and material of solar composition, C is dominated by C from the He intershell and the $^{12}$C/$^{13}$C ratio of the mixture will be relatively close to that in the outer He intershell. In RUN106, $^{12}$C/$^{13}$C =

6.6 is predicted for the He intershell, which gives $^{12}C/^{13}C = 8.2$ in the mixture. The predicted $^{26}Al/^{27}Al$ is $4.4 \times 10^{-6}$, much lower than observed in the grains. A higher $^{26}Al/^{27}Al$ and slightly lower $^{12}C/^{13}C$ could be achieved if the parent stars experienced cool bottom processing during their AGB phase (Nollett et al. 2003). $^{14}N/^{15}N$ is predicted to be 790.

In conclusion, born-again AGB stars appear to be stellar sites that provide a natural source of $^{32}S$ excesses via radioactive $^{32}Si$ decay along with low $^{12}C/^{13}C$ ratios, as observed in some AB grains. The models of Herwig et al. (2011) used here are 1D and spherically symmetric and it will be interesting to see whether full 3D models can quantitatively match the grain data in a self-consistent way. Also, future investigations are needed to evaluate the effect of GCE and heterogeneities in the ISM on S-isotopic compositions.

*Acknowledgements* – We thank Joachim Huth for the SEM analyses, Elmar Gröner for technical support on the NanoSIMS, and Alexander Heger for providing detailed SN data on www.nucleosynthesis.org. WF acknowledges support by Grant-in-Aid for Research Fellows of the Japan Society for the Promotion of Science (JSPS). EZ was supported by NASA grant NNX11AH14G. NuGrid acknowledges support from NSF grants PHY 02-16783 and PHY 09-22648 (Joint Institute for Nuclear Astrophysics, JINA) and EU MIRG-CT-2006-046520. The continued work on codes and dissemination of data is made possible through funding from STFC and NWERS Discovery grant (FH, Canada), and an Ambizione grant of the SNSF (MP, Switzerland). MP also thank for support from EuroGENESIS. NuGrid data is served by Canfar/CADC. We thank an anonymous reviewer for helpful and constructive comments.

**FIGURE LEGENDS**

**Figure 1.** Silicon-isotopic compositions of presolar SiC AB grains identified in this work. The squares represent grains that have Si-isotopic compositions away from the MS line by more than 2σ and which were measured for S- and in part also for N-, Mg-Al-, and Ca-Ti-isotopic compositions. Other AB grains (diamonds) and the mainstream grains (circles) from this study are shown for reference.

**Figure 2.** Sulfur-isotopic compositions of 34 SiC AB grains. Three grains (AB21, 24 and 40) show large $^{32}$S excesses of >100 ‰. Also shown are X and C grains with significant S isotope anomalies from previous studies (Gyngard et al. 2010a; Hoppe et al. 2012; Xu et al. 2012). The predicted S-isotopic compositions of the different zones in a 15 M$_\odot$ SNII (Rauscher et al. 2002) are shown by crosses and arrows for comparison.

**Figure 3.** Mass fractions of $^{12}$C, $^{26}$Al, $^{28}$Si, $^{32}$Si, $^{32}$S and $^{44}$Ti (upper panel), and solar-normalized ratios of $^{29,30}$Si/$^{28}$Si and $^{33,34}$S/$^{32}$S (lower panel, left scale), and the $^{32}$Si/$^{28}$Si ratio (lower panel, right scale) in the interior of a 15M$_\odot$ SNII model (Rauscher et al. 2002). The different SN zones are shown at the top of this figure.

**Figure 4.** Mass fractions of $^4$He, and C and Si isotopes in the He intershell of a born-again AGB star predicted by the model (RUN106) of Herwig et al. (2011).

**Table 1.** Isotopic compositions and trace element abundances of 34 SiC AB grains.

| Grain | $^{12}C/^{13}C$ | $^{14}N/^{15}N$ | $^{26}Al/^{27}Al$ $(10^{-3})$ | [Al][1] | $\delta^{29}Si$ (‰) | $\delta^{30}Si$ (‰) | $\delta^{33}S$ (‰) | $\delta^{34}S$ (‰) | [S][1] | $^{44}Ti/^{48}Ti$ $(10^{-2})$ | [Ti][1] |
|---|---|---|---|---|---|---|---|---|---|---|---|
| *AB grains with $^{32}S$ excesses* | | | | | | | | | | | |
| M7_AB21 | 5.71 ± 0.07 | | | | 2 ± 14 | 55 ± 17 | −600 ± 230 | −340 ± 130 | 0.39 | <26 | 0.06 |
| M7_AB24 | 7.44 ± 0.06 | | 3.28 ± 0.96 | 7.8 | 55 ± 14 | 100 ± 17 | −310 ± 140 | −131 ± 67 | 0.78 | | |
| M7_AB40 | 2.33 ± 0.01 | | 5.3 ± 1.2 | 17.3 | 37 ± 4 | 89 ± 4 | −320 ± 200 | −264 ± 90 | 0.12 | 0.33 ± 0.36 | 0.15 |
| *Other AB grains* | | | | | | | | | | | |
| M7_AB1 | 6.05 ± 0.02 | 3550 ± 570 | 1.6 ± 1.5 | 9.1 | 37 ± 6 | −47 ± 7 | 630 ± 270 | 34 ± 91 | 0.11 | 3.2 ± 1.7 | 0.68 |
| M7_AB2 | 3.18 ± 0.01 | | 2.72 ± 0.76 | 16.7 | 45 ± 3 | 21 ± 4 | 57 ± 82 | −13 ± 34 | 0.37 | | |
| M7_AB6 | 3.89 ± 0.02 | | 1.91 ± 0.42 | 7.6 | 6 ± 2 | 4 ± 3 | −47 ± 116 | 25 ± 52 | 1.41 | | |
| M7_AB7 | 2.10 ± 0.01 | | | | −37 ± 5 | −39 ± 6 | 213 ± 94 | 14 ± 37 | 0.90 | 0.02 ± 0.13 | 1.59 |
| M7_AB8 | 4.65 ± 0.03 | | 3.4 ± 1.3 | 2.6 | 17 ± 7 | −8 ± 8 | −270 ± 170 | −22 ± 84 | 0.38 | | |
| M7_AB10 | 5.27 ± 0.09 | | 0.19 ± 0.11 | 50.0 | 13 ± 9 | −11 ± 10 | −98 ± 190 | -8 ± 85 | 3.06 | | |
| M7_AB11 | 4.70 ± 0.03 | | | | −23 ± 5 | −21 ± 7 | −6 ± 140 | 7 ± 60 | 1.05 | | |
| M7_AB12 | 5.31 ± 0.06 | | 0.1 ± 2.7 | 12.9 | 60 ± 7 | 36 ± 8 | −1 ± 260 | −287 ± 92 | 0.90 | | |
| M7_AB13 | 5.08 ± 0.04 | | | | −21 ± 10 | −33 ± 12 | −68 ± 42 | 21 ± 19 | 14.2 | | |
| M7_AB14 | 4.83 ± 0.07 | | 0.52 ± 0.88 | 5.3 | −19 ± 13 | −37 ± 15 | 89 ± 212 | 43 ± 89 | 1.86 | | |
| M7_AB15 | 4.59 ± 0.07 | | | | −4 ± 7 | −9 ± 8 | −74 ± 226 | -89 ± 95 | 2.15 | | |
| M7_AB16 | 2.93 ± 0.03 | | 3.80 ± 0.74 | 5.3 | 49 ± 10 | 23 ± 12 | 180 ± 230 | −204 ± 79 | 0.95 | <0.72 | 0.85 |
| M7_AB17 | 8.93 ± 0.15 | | | | 61 ± 14 | 22 ± 17 | 33 ± 33 | −20 ± 14 | 73.9 | | |
| M7_AB18 | 4.46 ± 0.06 | | 1.4 ± 1.3 | 4.0 | 134 ± 14 | 79 ± 17 | 78 ± 159 | 64 ± 68 | 3.26 | | |
| M7_AB19 | 7.97 ± 0.04 | | | | 24 ± 6 | 48 ± 8 | 170 ± 150 | −35 ± 58 | 0.42 | | |
| M7_AB20 | 2.62 ± 0.03 | | 10.2 ± 2.0 | 3.9 | −39 ± 13 | 22 ± 16 | −400 ± 170 | −62 ± 91 | 1.29 | <4.0 | 0.47 |
| M7_AB22 | 8.34 ± 0.17 | | <2.5 | 27.1 | −24 ± 6 | 17 ± 8 | −210 ± 230 | 140 ± 120 | 1.72 | | |
| M7_AB25 | 3.57 ± 0.01 | 63 ± 1 | 2.4 ± 4.5 | 46.2 | 34 ± 6 | 60 ± 7 | −68 ± 97 | 27 ± 44 | 0.97 | | |
| M7_AB26 | 7.52 ± 0.02 | | 0.55 ± 0.80 | 2.0 | −27 ± 13 | 43 ± 16 | 134 ± 59 | 16 ± 24 | 0.77 | | |
| M7_AB27 | 1.91 ± 0.01 | | 9.0 ± 9.0 | 7.2 | 9 ± 19 | 103 ± 24 | 110 ± 240 | −228 ± 84 | 0.60 | 0.5 ± 1.0 | 0.61 |
| M7_AB29 | 5.83 ± 0.08 | | | | −28 ± 27 | 124 ± 36 | 470 ± 310 | −35 ± 106 | 1.05 | | |
| M7_AB30 | 7.81 ± 0.03 | | <2.8 | 6.7 | 8 ± 4 | 40 ± 5 | −79 ± 143 | 152 ± 68 | 0.21 | <7.5 | 0.68 |
| M7_AB31 | 8.07 ± 0.04 | | | | 19 ± 9 | 81 ± 11 | −29 ± 129 | 90 ± 58 | 0.31 | | |
| M7_AB33 | 6.55 ± 0.05 | | | | 63 ± 17 | 197 ± 22 | −89 ± 115 | 44 ± 53 | 1.14 | | |
| M7_AB34 | 5.49 ± 0.10 | | 56.5 ± 8.2 | 12.2 | −5 ± 10 | 86 ± 12 | 110 ± 120 | −57 ± 49 | 8.78 | | |
| M7_AB35 | 4.24 ± 0.02 | | | | 81 ± 3 | 98 ± 4 | 200 ± 150 | 82 ± 65 | 0.28 | 0.39 ± 0.31 | 0.13 |
| M7_AB36 | 2.85 ± 0.01 | | 13.5 ± 1.6 | 35.4 | −20 ± 5 | 50 ± 7 | −340 ± 150 | −4 ± 79 | 0.14 | <0.20 | 1.64 |
| M7_AB37 | 2.19 ± 0.01 | | | | 4 ± 4 | 55 ± 5 | 56 ± 69 | −87 ± 27 | 1.57 | 1.2 ± 0.7 | 0.31 |
| M7_AB38 | 1.85 ± 0.00 | | 2.08 ± 0.62 | 14.8 | 8 ± 4 | 56 ± 5 | 116 ± 74 | −11 ± 30 | 0.34 | | |
| M7_AB39 | 6.03 ± 0.02 | | | | 10 ± 5 | 82 ± 7 | 1 ± 130 | 80 ± 57 | 0.24 | | |
| M7_AB41 | 9.90 ± 0.06 | | | | −26 ± 3 | 31 ± 3 | 120 ± 100 | 14 ± 42 | 0.90 | | |
| Average[2] AB21, 24, 40 | 3.09 ± 0.01 | | 4.65 ± 0.86 | | 36 ± 3 | 88 ± 4 | −360 ± 100 | −202 ± 49 | 0.43 | 0.24 ± 0.42 | |
| SN model A | 17 | | 3.97 | | 30 | 90 | −253 | −234 | | 0.06 | |
| SN model B | $2.2 \times 10^5$ | | 4.1 | | 33 | 91 | −250 | −250 | | 4.65 | |
| Born-again AGB model | 8.2 | 790 | 0.0044 | | 37 | 161 | | | | | |

[1]Weight percent.
[2]Averages are calculated by adding counts from individual measurements. Errors are 1σ and are based on counting statistics only.

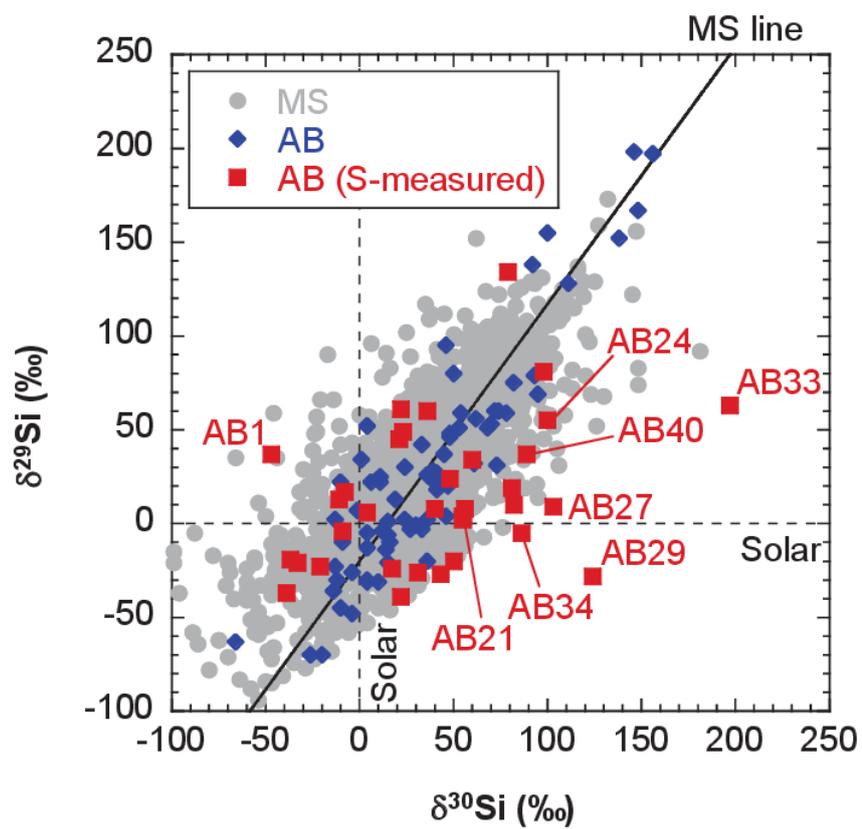

Figure 1

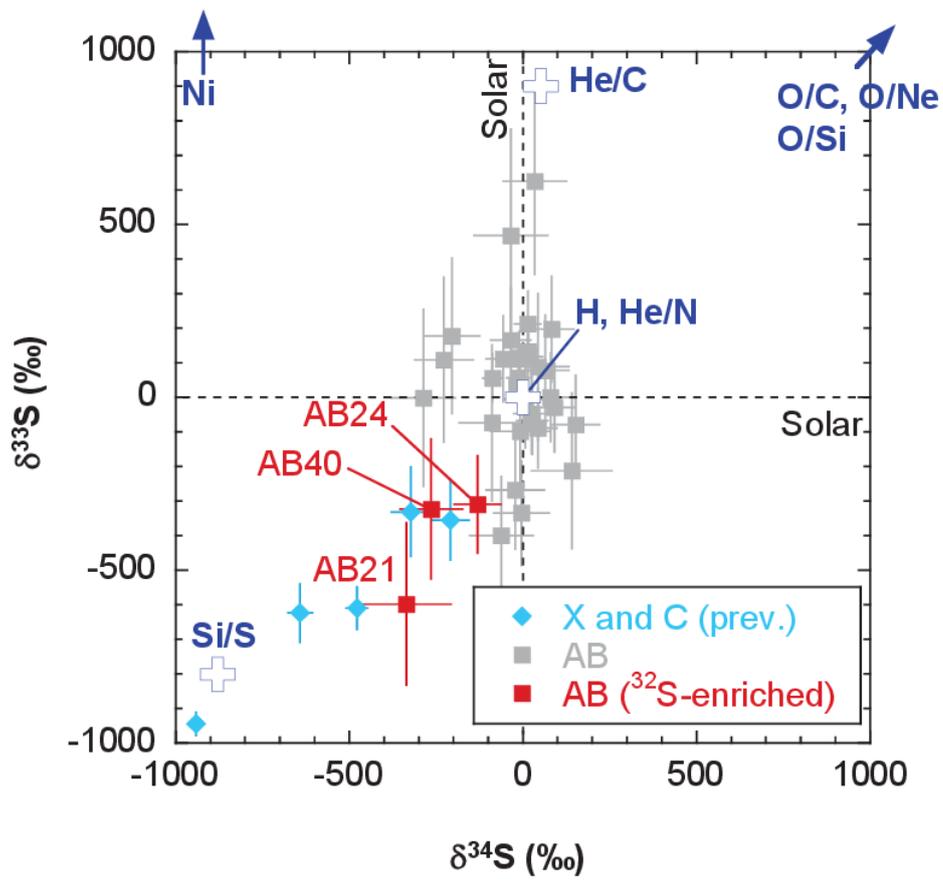

Figure 2

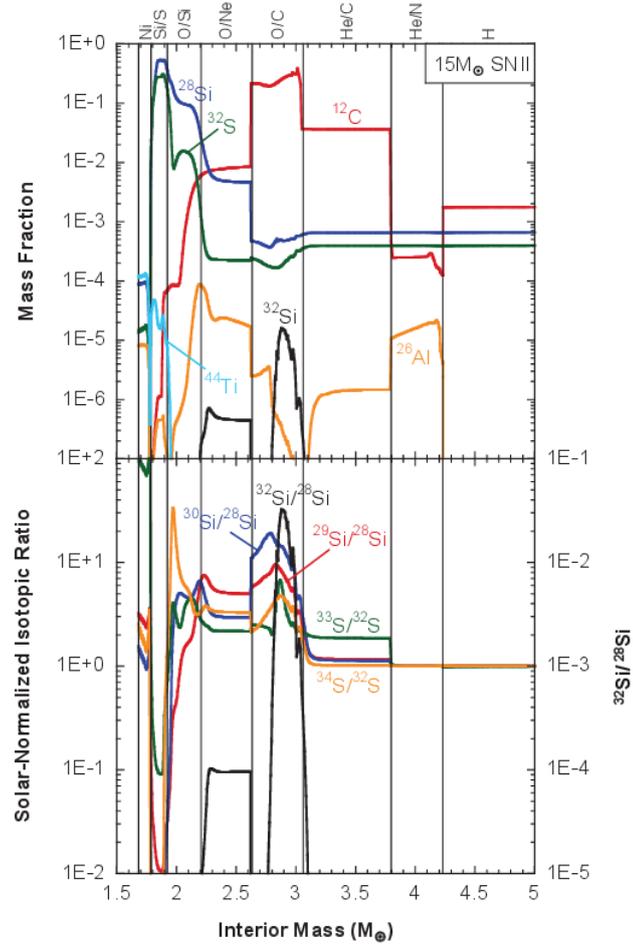

Figure 3

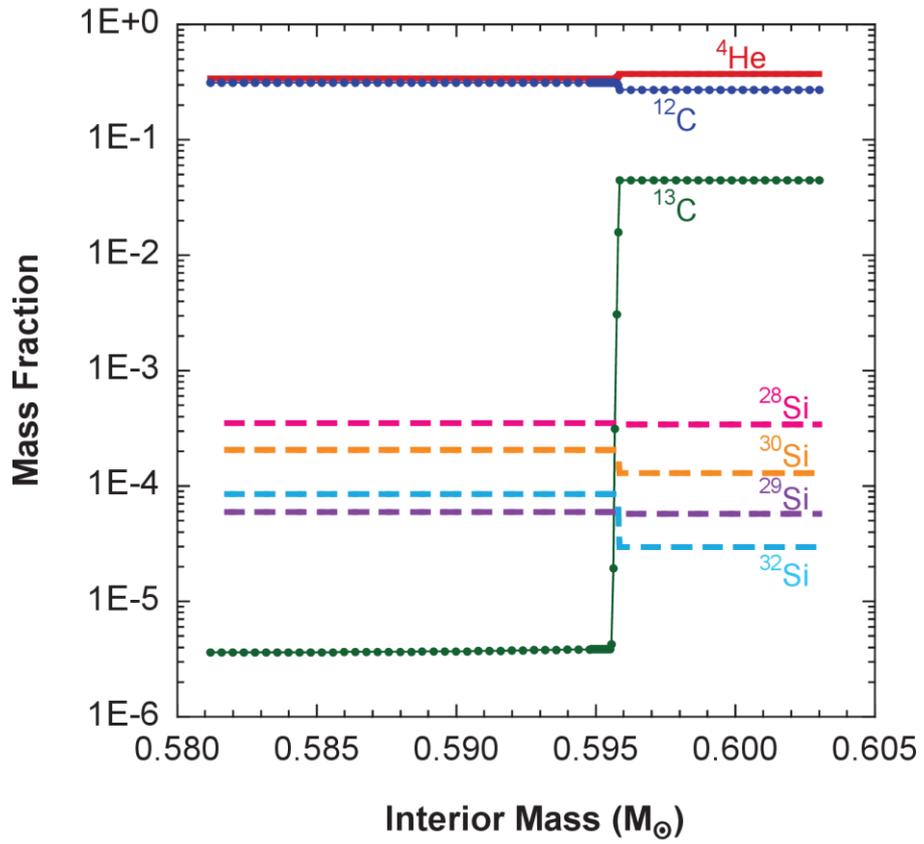

Figure 4